\documentclass{article}

\usepackage{arxiv}

\usepackage[utf8]{inputenc} 
\usepackage[T1]{fontenc}    
\usepackage{graphicx}
\usepackage{hyperref}
\usepackage{natbib}
\hypersetup{colorlinks,allcolors=black}
\usepackage{comment}
\usepackage{amsmath}	
\usepackage{amssymb}	
\usepackage{subcaption}
\usepackage{url}            
\usepackage{booktabs}       
\usepackage{amsfonts}       
\usepackage{nicefrac}       
\usepackage{microtype}      
\usepackage{lipsum}

\def\fl{{\it Fermi}-LAT }

\def\beq{\begin{equation}} 
\def\eeq{\end{equation}} 
\def\beqar{\begin{eqnarray}} 
\def\eeqar{\end{eqnarray}}

\def\pfrac#1#2{\left( \frac{#1}{#2} \right)}

\def \nn{\nonumber}

\def \l({\left(}
\def \r){\right)}

\def\pfrac#1#2{\left( \frac{#1}{#2} \right)}

\def\Gm{\Gamma}

\def\f#1{f_{\rm #1}}
\def\t#1{t_{\rm #1}}
\def\dinv{{\rm day^{-1}}}

\title{Estimating longterm power spectral densities 
in AGN from simulations}

\author{
  Nachiketa ~Chakraborty\thanks{Guest Researcher at Max-Planck-Institut f\"ur Kernphysik} \\
  Data Assimilation Research Centre, Department. of Meteorology \\ University of Reading \\
  Reading RG66BG, UK\\
  \texttt{n.chakraborty@reading.ac.uk} \\
  \texttt{cnachi@mpi-hd.mpg.de} \\
   \And
 Frank M. Rieger \\
  Max-Planck-Institut f\"ur Kernphysik \\  
  69117 Heidelberg, Germany\\
  \texttt{frank.rieger@mpi-hd.mpg.de} \\
}

\begin{document}
\maketitle

\begin{abstract}
The power spectral density (PSD) represents a key property quantifying the stochastic 
or random noise type fluctuations in variable sources like Active Galactic Nuclei (AGN). 
In recent years, estimates of the PSD have been refined by improvements in both, the 
quality of observed lightcurves and modeling them with simulations. This has aided in
quantifying the variability including evaluating the significance of quasi-periodic 
oscillations. A central assumption in making such estimates is that of weak 
non-stationarity. This is violated for sources with a power-law PSD index steeper 
than one as the integral power diverges. As a consequence, estimates of the flux 
probability density function (PDF) and PSD are interlinked. In general, for evaluating 
parameters of both properties from lightcurves, one cannot avoid a multi-dimensional, 
multi-parameter model which is complex and computationally expensive, as well as harder 
to constrain and interpret. However, if we only wish to compute the PSD index as is 
often the case, we can use a simpler model. We explore a bending power-law model 
instead of a simple power-law as input to time-series simulations to test the quality 
of reconstruction. Examining the longterm variability of the classical blazar Mrk 421, 
extending to multiple years as is typical of \fl or Swift-BAT lightcurves, we find 
that a transition from pink (PSD index one) to white noise at a characteristic 
timescale, $t_b \sim 500-1000$ years, comparable to the viscous timescale at the disk 
truncation radius, seems to provide a good model for simulations. This is both 
a physically motivated as well as a computationally efficient model that can be used 
to compute the PSD index. 
\end{abstract}


\section{Introduction}
A key area of time domain astronomy involves the modeling of noisy time-series or
lightcurves. This usually requires the shape of the noise spectrum or the power 
spectral density (PSD) as input. 
Several estimation problems of temporal features depend upon the type and level of
noise that acts as background. For example, estimation of the significance of
quasi-periodic oscillations depends on the assumed level of noise which in turn
depends upon the PSD \citep[e.g.,][]{2015ApJ...813L..41A,vau2016,cov2019,ait2020}.
Estimation of variability properties also depends upon the probability distribution
function (PDF) from which fluxes are drawn. Indeed, the PDF and the PSD are the two 
key properties that are central to quantifying statistically, the variability of a 
source with stochastic processes. They are the primary observables upon which other
observables that allow us to probe the physical processes driving variability,
depend on. Consequently, they both are also key inputs to simulation methods used 
to statistically characterise the variability properties. Such simulation techniques
are based on spectral density estimation, a fundamental aspect of time domain
astrophysics that has seen development at different stages in the last few decades 
\citep{TK:95, 2013arXiv1309.6435V}. 

In principle, the PDF and the PSD are independent inputs or observables. The 
PSD represents the distribution of power at different timescales. It carries
information about the strength of the physical processes driving variability
at particular timescales \citep[e.g.,][for a recent discussion]{rie2019}. 
Often the functional form of the PSD can be approximated by a power-law that 
may include a characteristic break separating regions of different indices
\citep[e.g.,][]{2002MNRAS.332..231U}. The PDF, on the other hand, represents 
the probability distribution of the flux itself and encodes the form or class 
of the physical process. For instance, the PDF can discriminate between 
additive and multiplicative processes depending on whether it is normal or 
lognormal \citep{UMV:2005}. However, these observables as inputs to simulation 
both impact on the estimation of variability. And through this degeneracy of 
effects, they interact with each other. For instance, the estimates of the PSD 
shape with the Timmer-Koenig (hereafter TK95) method relies on the assumption 
that the PDF is Gaussian or normal. If it fails, the estimates are no longer 
reliable as shown in e.g., \citet{2003MNRAS.345.1271V} and \citet{2019MNRAS.489.2117M}. 
This is because of the divergence of total power leading to weak non-stationarity 
as explained in e.g., \citet{2003MNRAS.345.1271V} and \citet{UMV:2005}.

A variety of PSD analyses have been performed in recent times for the longterm 
lightcurves of gamma-ray emitting Active Galactic Nuclei (AGN), particularly 
in the high-energy \fl domain. In the majority of cases, the inferred PSDs 
appear compatible with power laws $P(f) \propto f^{-\Gamma}$ with $\Gamma \sim 0.8
-1.6$ \citep[e.g.,][]{abdo2010,ack2011,sob14,max2014,hess_pks_longterm,kush2017,
bhat2020,ait2020,2020MNRAS.494.3432G}

In this paper, we wish to have a revised look at the problem of PSD estimation. 
Due to the presence of weak non-stationarity for PSD indices steeper than one, 
we need to find ways to prevent the total power at low frequencies from diverging.
One convenient way to do this, is to introduce a break, $\f{break}$ or a bend to 
transition from coloured noise, appropriate for the observed variations, to white 
noise that we will eventually be reduced to at long enough timescales. This is 
because, for any real object the variability power spectra must converge. We 
cannot have an infinitely large quantity of power in variations at any timescale 
including the longest ones.
While this is, a priori a mathematical technique to avoid divergence, the presence 
of breaks in the PSD is also motivated physically. Thus, in this paper we explore
the mathematical effect of inserting a break or a bend in determining the PSD 
using simulation, taking multi-year lightcurves of the blazar Mrk 421 as example. 
We also provide a physical motivation for the presence of breaks, and compare it 
to those that work for simulations. 

Another approach is to try and model accurately the shape of the true PDF. This 
is certainly a worthy approach and the sophisticated simulation technique in 
\citet{2013MNRAS.433..907E} would be the framework to adopt (we will refer to 
it as DE13 henceforth). However, there are at least two challenges. First, as 
we test more complex models than a simple power-law PSD, we have several free 
parameters (e.g., four for a broken power-law) and need long lightcurves with a 
fine cadence to constrain them adequately. Secondly, and perhaps more crucially, 
approaches like DE13 are computationally very expensive. This makes it rather 
challenging to adopt these without significant modifications to speed up the 
algorithms, for detecting and significance testing temporal features in big data 
sets.

In this study, we focus on the long term variability behaviour and not on short 
outbursts or flares. We are interested in studying the transition regime(s) at
timescales longer than those for fast variability (occurring on scales of minutes).
The flares represent sharp and transient changes in variability properties 
potentially leading to a departure from the underlying distribution or flux
probability distribution function (PDF) of the default, persistent mechanism. 

From \citet{2019MNRAS.489.2117M} we know that for power-law PSD indices steeper 
than pink noise ($\Gamma>1$), there is weak non-stationarity. This affects the 
time-series simulations in that the PDF does not remain fixed for the entire 
ensemble. 
This is true independent of red noise leakage corrections as it has to do with 
divergence of power at longer timescales. Here we explore a way to address this 
by introducing a break or a bend in the PSD. 
The position of this break then becomes crucial and could either come directly 
from physical models (e.g., viscosity timescale) or in absence of this, from 
methodologically testing what explains best the observations. 
In the present paper we use simulated observations of known PSD indices to explore 
the relation between a physical break and a methodological break. We find that 
the two converge for long term variations akin to multi-year \fl or Swift-BAT 
lightcurves to within a factor of a few.


The paper is organised as follows. In section~\ref{sect:analytictoy}, we illustrate 
first with an analytical model and then with a physical model of accretion disk 
powered fluctuations, the effect of and need for the break in the PSD. This will 
then be tested with artificial lightcurves generated by the time-series simulations 
described in the next section~\ref{sect:method}. Section~\ref{sect:methodtests} 
explains the results of the tests evaluating the position of the break or bend on
reconstructing the "true" PSD from observed lightcurves, both using artificially 
generated power-laws as well as real observations of Mrk 421, a classical blazar
at X-ray and gamma-ray energies.
In the final section~\ref{sect:conc}, we present a discussion of the results and 
conclusions. 

\section{General Considerations}\label{sect:analytictoy}

\noindent

\begin{figure}
    \centering
    \includegraphics[width=0.45\textwidth]{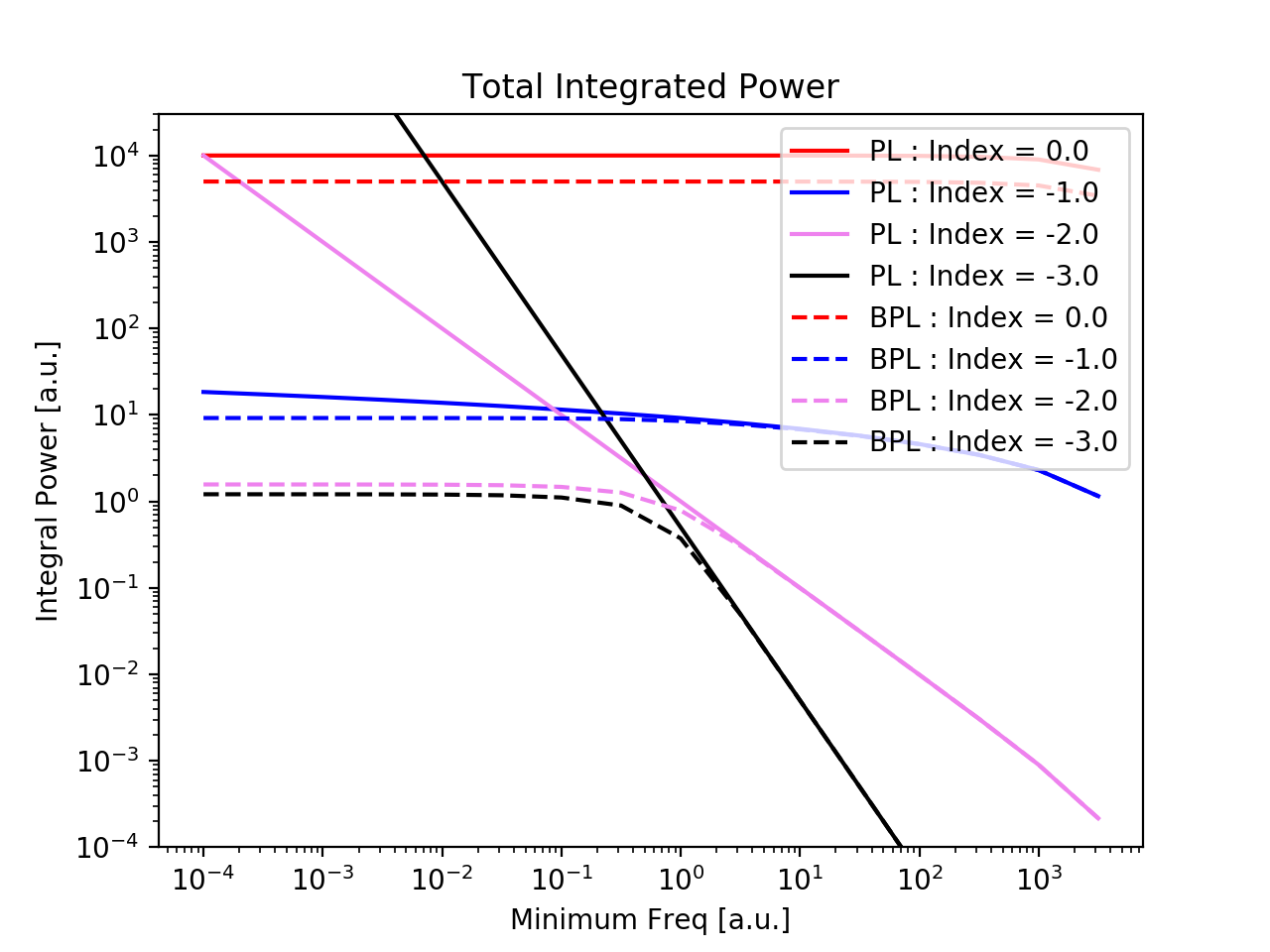}
    \caption{Figure illustrating the difference in total power between extending a power-law 
    model of indices [0,3] and a bending power law with a bend or break frequency at one. 
    Obviously, the effects are stronger for steeper indices. Therefore, the sensitivity to the position of the break $\f{break}$ is also greater for sharper transitions.}
    \label{fig:BPLvsPLindices}
\end{figure}

As is evidenced by prior endeavours at PSD estimation and its applications
\citep[e.g.,][]{2003MNRAS.345.1271V,UMV:2005,2013arXiv1309.6435V,2019MNRAS.489.2117M}, 
there are several effects at play. In order to disentangle some of these effects 
and build an intuitive understanding, we employ two toy models.

\subsection{PSD characterizations}
We take simple power law (PL) and broken or bending power law (BPL) models to 
investigate the power as a function of different parameters like the break 
frequency, $\f{b}$, the power law index, $\Gamma$, and the minimum frequency,
$\f{min}$ which is associated with the longest timescales. 
The PL and BPL models are formally given by,
\begin{equation}
\label{eqn:PLmodel}
    P(f) = A\ \pfrac{f}{\f{b}}^{-\Gamma}
\end{equation}
and,
\begin{equation}
\label{eqn:BPLmodel}
    P(f) = \frac{A}{1 + \pfrac{f}{\f{b}}^{\Gamma}}\,.
\end{equation}
For the chosen BPL model the transition from a power-law type shape with 
index $\Gamma$ to white noise (of index zero) is a smooth one. This is 
also the mathematical form appropriate for an Ornstein-Uhlenbeck (OU) 
process which is a popular model for noisy lightcurves in quasars
\citep[e.g.,][]{2009ApJ...698..895K}. 
The OU process has a constant diffusion coefficient and is key to a 
mathematical description of Brownian and Johnson noise 
\citep[see][for a review]{1996AmJPh..64..225G}. On the other hand,
a simple broken power-law (BrPL) model, as given by 
\beqar
\label{eqn:BrPLmodel}
    P(f) &=& A \ \ \ \ \ \ \ \ \ \ \ \ \ \ \ \  \ \   (f<\f{b}) \nn \\
         &=& A\ \pfrac{f}{\f{b}}^{-\Gamma} \ \ \ \ \ \  (f\geq\f{b})\,,
\eeqar
is less smooth and in that way less physical.

We then evaluate the total power in a given frequency band, $[\f{min},\f{max}]$ 
as a function of $\f{min}$ for different PL and BPL models. This provides a 
quantitative understanding of the amount of power below the bend for different 
values of index $\Gamma$, which can be transferred to higher frequencies above 
the bend. 
As shown in figure~\ref{fig:BPLvsPLindices}, it is clear that as $\f{min} 
\rightarrow 0$ the total power converges or asymptotes to a fixed value for 
the BPL models rather than diverge as for the PL models. This is because the 
noise properties transition to white noise, and below a certain frequency 
(or above the equivalent timescale) this white noise dominated regime will not 
have a significant amount of additional power by extending the bandwidth to 
lower frequencies. 

For a finite bandwidth, naturally the integral power is finite and does not 
diverge. 
It is obvious, however, that the difference in total power between each pair 
of PL and BPL with the same $\Gamma$ increases with steeper indices, and 
becomes quite significant as we go to indices greater than one. This marked 
difference is essentially responsible for making the PSD estimation more 
challenging beyond this value. It also illustrates that the absolute difference 
in total power depends upon the value of the break frequency. 
The lower the break frequency, the greater is the power, both differential and 
integral at longer timescales. As we will see in the subsequent sections, for 
the simulated ensembles there appears to be an optimal choice for the break 
frequency that fixes the mathematical issue of weak non-stationarity / power 
divergence and satisfies the physical constraint related to accretion disk 
fluctuations.

\subsection{Mathematical motivation : Extending PL validity range 
vs minimising leakage}\label{sect:breakplvsleak}
In order to counter weak non-stationarity in simulated lightcurves, we introduce 
a bend or break transitioning to white noise at low frequencies, thus avoiding 
the divergence of total power. As stated before, in introducing a break or bend 
$\f{b}$ in a power-law there are two key factors to take into account. The 
first is that this artificial break, $\f{b}$ should not introduce any spurious 
effects within the observed Fourier bandwidth, $[\f{min},\f{max}]$. In other words, 
if we want to test PL models for the observed variability, we must ensure that 
curvature from a break close to $\f{min}$ should not cause simulations to deviate 
from this. Hence, we position the break outside the observed bandwidth at lower 
frequencies. Intuitively, one might imagine that placing $\f{b}$ increasingly 
farther from the minimum observed Fourier frequency, $\f{min}$, should improve the
power-law behaviour within the observed band, $[\f{min},\f{max}]$ and hence the 
PSD estimation, until lowering the frequency further has no perceptible effects.
However, we find that this may not be the case and there seems to be an 
optimal/preferred position of $\f{b}$, below which the results not only do not
improve, but in fact, deteriorate. This is clearly observed in figure~\ref{fig:0p6501}, 
and is explained in the next sections. 
This is where the second factor comes in. We know that for any lightcurve, 
restricted to a limited bandwidth, there is red noise leakage, or transfer of 
power from frequencies lower than $\f{min}$ to the higher frequency end near, 
$\f{max}$. This produces a flattening of the PSD index at the higher frequency 
end \citep{2016ApJ...825...56Z}. Introducing a bend, transitioning to lower index 
values and realistically to white noise, will ameliorate this effect, as the 
leakage power transferred from lower frequencies will be reduced. This is 
clearly seen in the toy model example in figure~\ref{fig:BPLvsPLindices}, showing 
the total integral power within a certain bandwidth as a function of minimum 
frequency $\f{min}$ to a fixed maximum. 
Note that this minimum frequency is not necessarily the minimum of observed band. 
In general, this bandwidth is larger than the observed bandwidth. It is obvious 
that for a PL (solid), the integral power in any bandwidth is greater than for the 
BPL model (dashed).
Furthermore, the black curve in the right panel of figure~\ref{fig:BPLvsPLindices} 
shows the extent of leaked power that will be transferred from low to high frequencies. 

However, if $\f{b} \ll \f{min}$, then the PL extension continues to much lower 
frequencies below $\f{min}$, increasing the relative red noise leakage factor. This 
will cause the results to deteriorate. As we will see, this is not a significant 
concern, as the physically motivated break frequency as described in 
section~\ref{physics} are in a regime that prevent us from going to arbitrarily 
low values. These toy models are shown for index 1.0, with effects more significant 
at steeper indices as shown in figure~\ref{fig:BPLvsPLindices}. This will help to 
explain the results of PSD estimation with the $\fl$ and Swift-BAT simulations. 

\subsection{Physical Motivation}\label{physics}
It seems likely that the variability in AGN seen on long timescales (i.e. short 
temporal frequencies f=1/t) is influenced by changes in the accretion flow. 
Consider for example the fluctuating disk model \citep{lyu1997}, in which 
fluctuations of the disk parameters at some radius, assumed to occur on local 
viscous timescale $t_{\rm visc}$, produce variations in the accretion rate at 
smaller radii that can be of the power-law PSD type. If these are effectively
transmitted to the jet, power-law PSD type variability might occur 
\citep[e.g.,][for a recent review]{rie2019}. Breaks might then naturally be 
expected given the limited extent $r_d$ for a stable disk configuration. In 
terms of the $\alpha$-parameter and the disk scale height to radius ratio $(h/r)$, 
the viscous timescale can be expressed as
\beq
 t_{\rm visc}(r) = \frac{1}{\alpha}\left(\frac{r}{h}\right)^2 
 \left(\frac{r}{r_g}\right)^{3/2} \frac{r_g}{c}\,.
\eeq For typical numbers, i.e. $\alpha=0.1$, $h/r = 0.1$ and $r = r_d \sim 10^3 
r_g$, $r_g = G M_{\rm BH}/c^2$ \citep{goo2003}, this would then yield a timescale 
at which a break might be expected of $t_{\rm b}=t_{\rm visc}(r_d) \sim 10^3
\,(M_{\rm BH}/10^8 M_{\odot})$ yr, and hence a characteristic low-frequency 
break $f_{\rm b}=1/t_{\rm b}$.

\subsection{Real data}
AGN variability studies across the electromagnetic spectrum 
\citep[e.g.,][]{2003ApJ...593...96M,chat2011,2012A&A...540L...2I,hess_pks_longterm, 
2019ApJ...885...12R,2020MNRAS.494.3432G} have explored PSD breaks 
corresponding to characteristic timescales of physical processes. 
At gamma-ray energies, typically there are two types of breaks reported. 
One at longer timescales, possibly associated with dynamical processes 
such as accretion disk fluctuation, and the other at relatively shorter 
timescales, possibly associated with particle transport (either radiative 
cooling such as Compton losses or particle acceleration or escape) 
\citep[e.g.,][]{2019ApJ...885...12R}. At X-ray energies, there has been 
a longer history of estimating break frequencies. These are typically
associated with a characteristic physical timescales of the accretion 
disc \cite[e.g.,][]{2003ApJ...593...96M,papa2004}, such as the viscus time 
at the truncation radius, though other possibilities exist 
\citep{2012A&A...540L...2I}. 

In our present study we use the multiwavelength lightcurves of the classical
blazar Mrk 421 ($z=0.031$) as obtained from the published analysis in 
\citet{421logN}. These observations last from from 2009 to 2015 and were led 
by the HAGAR Telescope Array collaboration as a part of a regular monitoring 
campaign of Mrk 421. Long-term properties of the source were studied from radio 
to gamma-rays.
We select from amongst the different observed lightcurves those at high energies, 
i.e. \fl (Large Area Telescope) at GeV energies and Swift-BAT (Burst Alert 
Telescope, 15-150 keV) in hard X-rays, to avoid spurious effects related
to particle cooling at lower energies \citep[e.g.,][]{fink2014} which may occur 
in, e.g., the radio band.

\section{Methodological approach}\label{sect:method}
\begin{figure}
\begin{center}
\scalebox{0.85}{
	\includegraphics[width=0.5\textwidth]{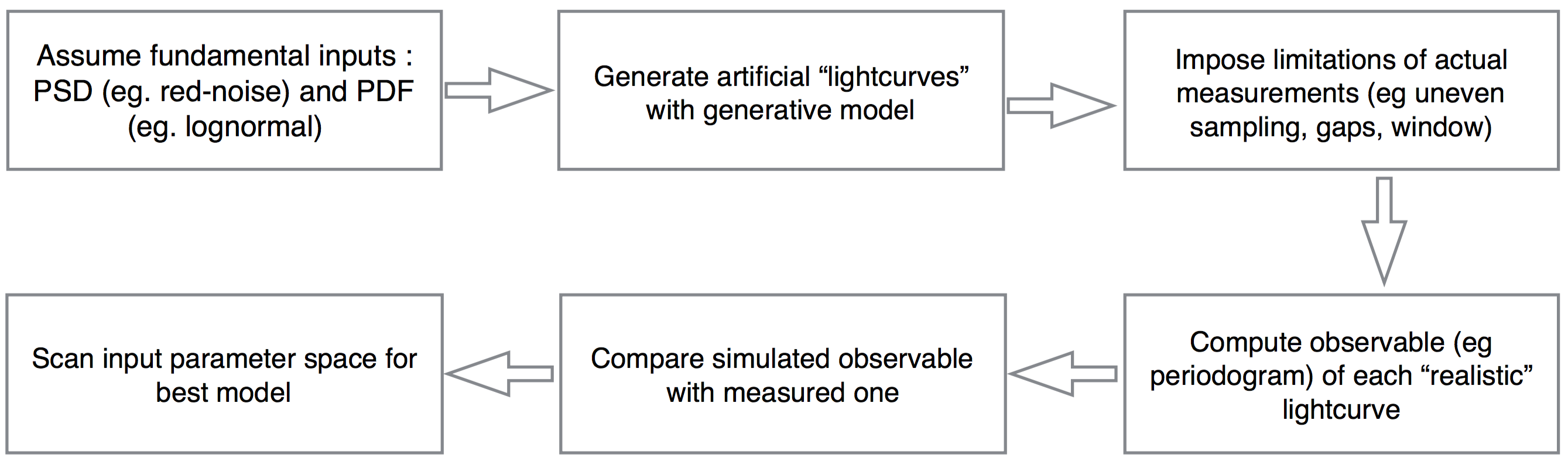}
}
\caption{The applied time-series method involves modifying the Timmer-Koenig 
method to account for lognormality and gaps (or Emmanoulopoulos method) to 
generate lightcurves of a given PSD and PDF. The observed cadence, i.e., gaps, 
sampling and duration of the window are imposed on the simulated lightcurves. 
The PSD and PDF parameters that best describe the observations are evaluated 
by a likelihood analysis.}
\label{fig:variabilityestimator}
\end{center}
\end{figure}

The time-series approach chosen here is outlined in the flowchart in 
figure~\ref{fig:variabilityestimator}. The steps in the method are as 
follows 

\begin{itemize}
\item {\bf Input "model"}: For variability studies, one can choose a theoretical, 
mathematical or phenomenological model and produce a lightcurve as input for the 
time-series simulation method in the next steps. 
\item {\bf Simulate ensemble}: Artificial lightcurves are generated with what 
is essentially a TK type \citep{TK:95} method. Thus, by default we have 
lightcurves with a Gaussian PDF and a power-law PSD. We modify this to allow for 
log-normal PDFs using the exponential transform when needed \citep[cf.][]{UMV:2005} 
as in \citet{2020Galax...8....7C}, and use a bending power-law model of the form 
eq.~\ref{eqn:BPLmodel}. For this we use the $BendingPL$ and 
$Simulate\_TK\_Lightcurve$ modules from \citet{2015arXiv150306676C}.
\item {\bf Impose observational cadence}: The artificial lightcurves in the 
previous step would be "ideal" in the sense of being continuous and long, with 
fine sampling and no gaps. However, for the ensemble we impose the actual 
observational window, gaps and time sampling, matching precisely the observed 
times. The resulting lightcurves are "realistic" simulations in that they factor 
in observational effects. 
\item {\bf Compute "observed" properties}: The "realistic" simulated lightcurves 
fold in these effects and therefore are accurate and precise representations of 
observational cadence. The computed variability is a convolution of physical 
processes and observational cadence effects and can be compared to the observed 
variability. 
\item {\bf Scan model parameter space}: Having computed variability for a particular 
set of model parameters, we repeat the aforementioned procedure scanning the parameter 
space of the chosen input model. 
\item {\bf Model selection / hypothesis testing }: To determine which model 
parameters best describe the observed variability properties we perform a 
chi-squared test.
\end{itemize}

\section{Methodological tests of PSD estimates}
\label{sect:methodtests}
\begin{figure}
\begin{subfigure}{.5\textwidth}
  \centering
  \includegraphics[width=.44\linewidth]{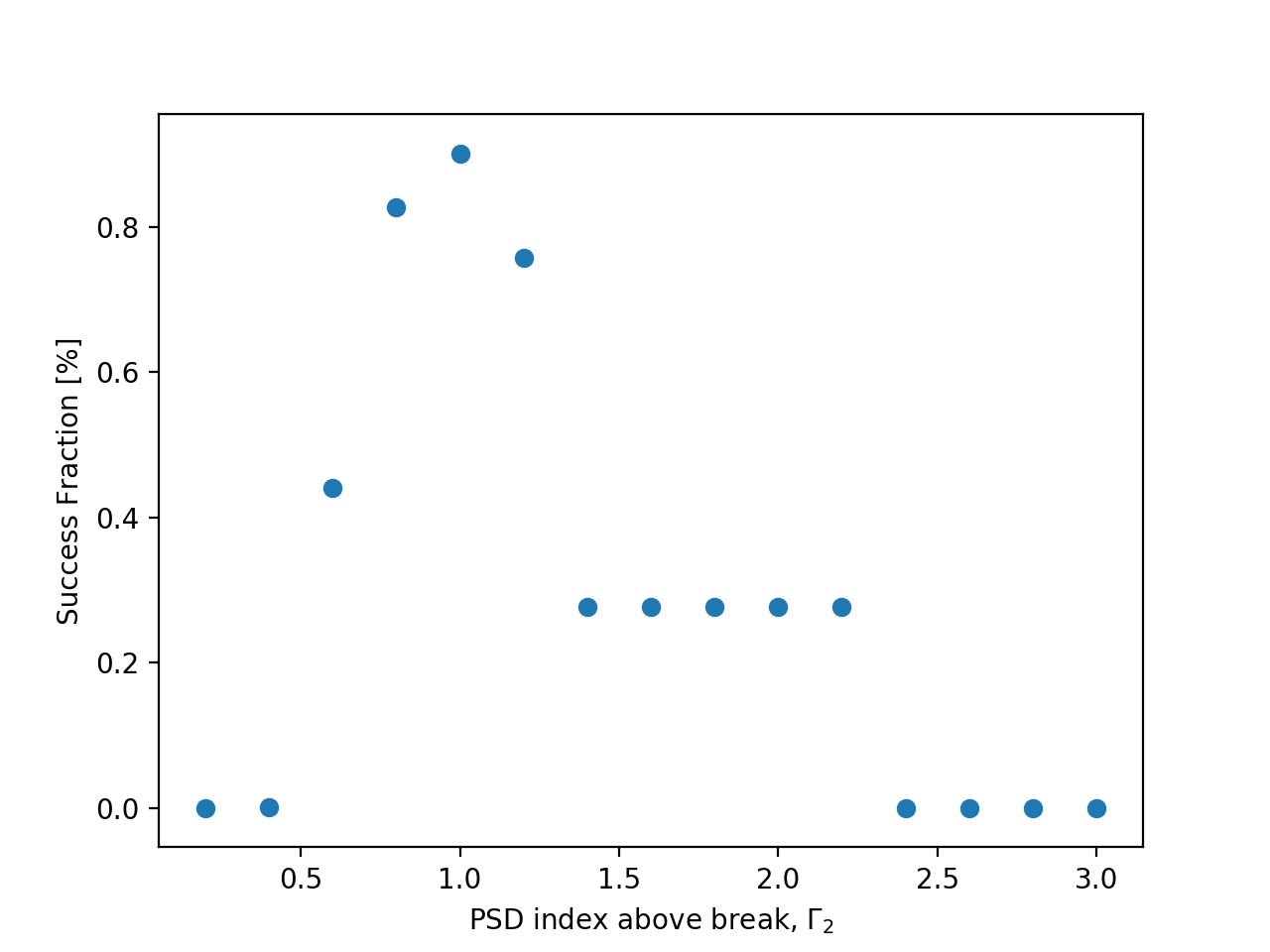}
  \includegraphics[width=0.44\textwidth]{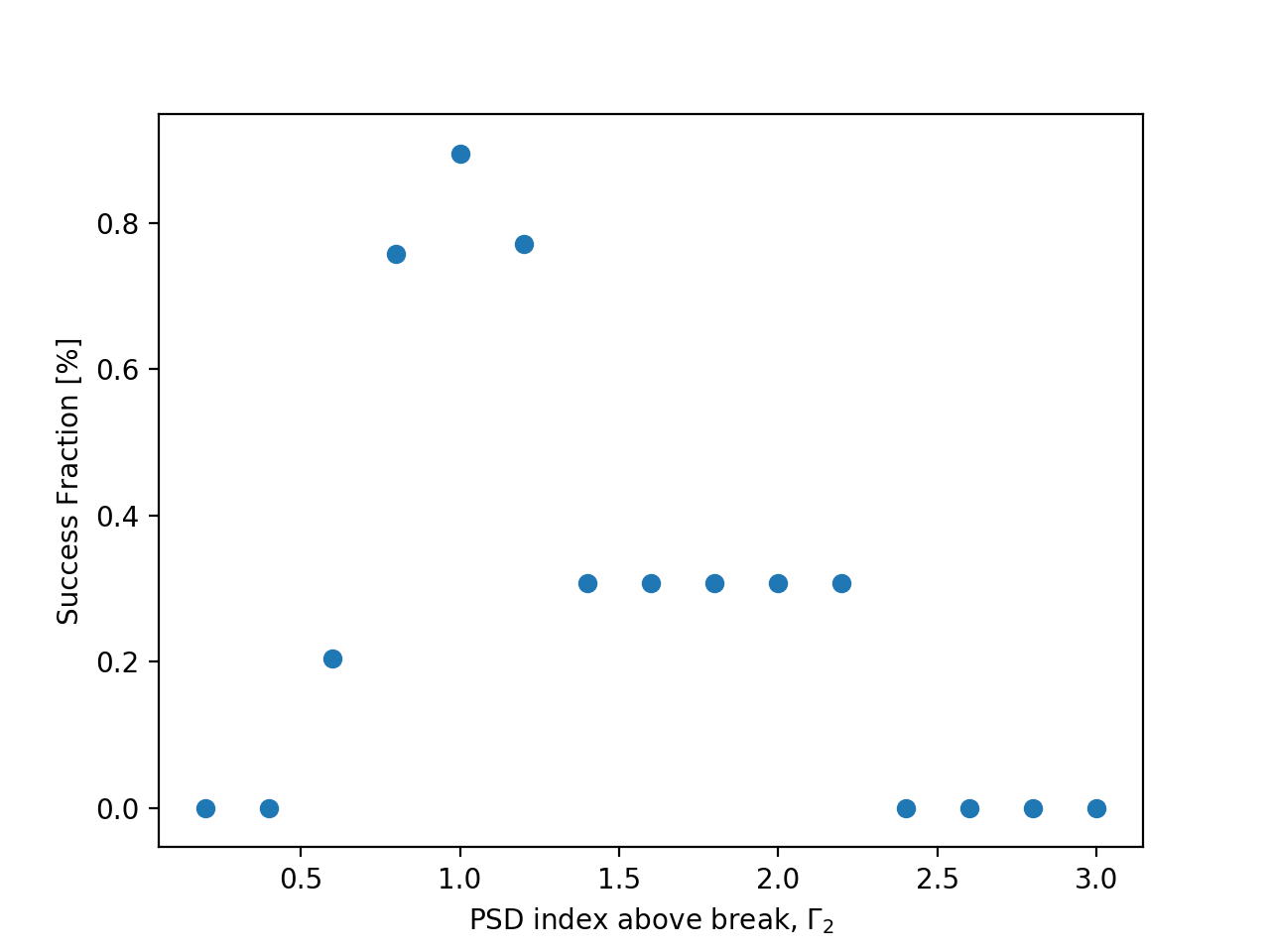}
  \caption{Pink noise or $\Gamma = 1.0$}
  \label{fig:BPLvsBrPLPSD1}
\end{subfigure}
\begin{subfigure}{.5\textwidth}
  \centering
  \includegraphics[width=.44\linewidth]{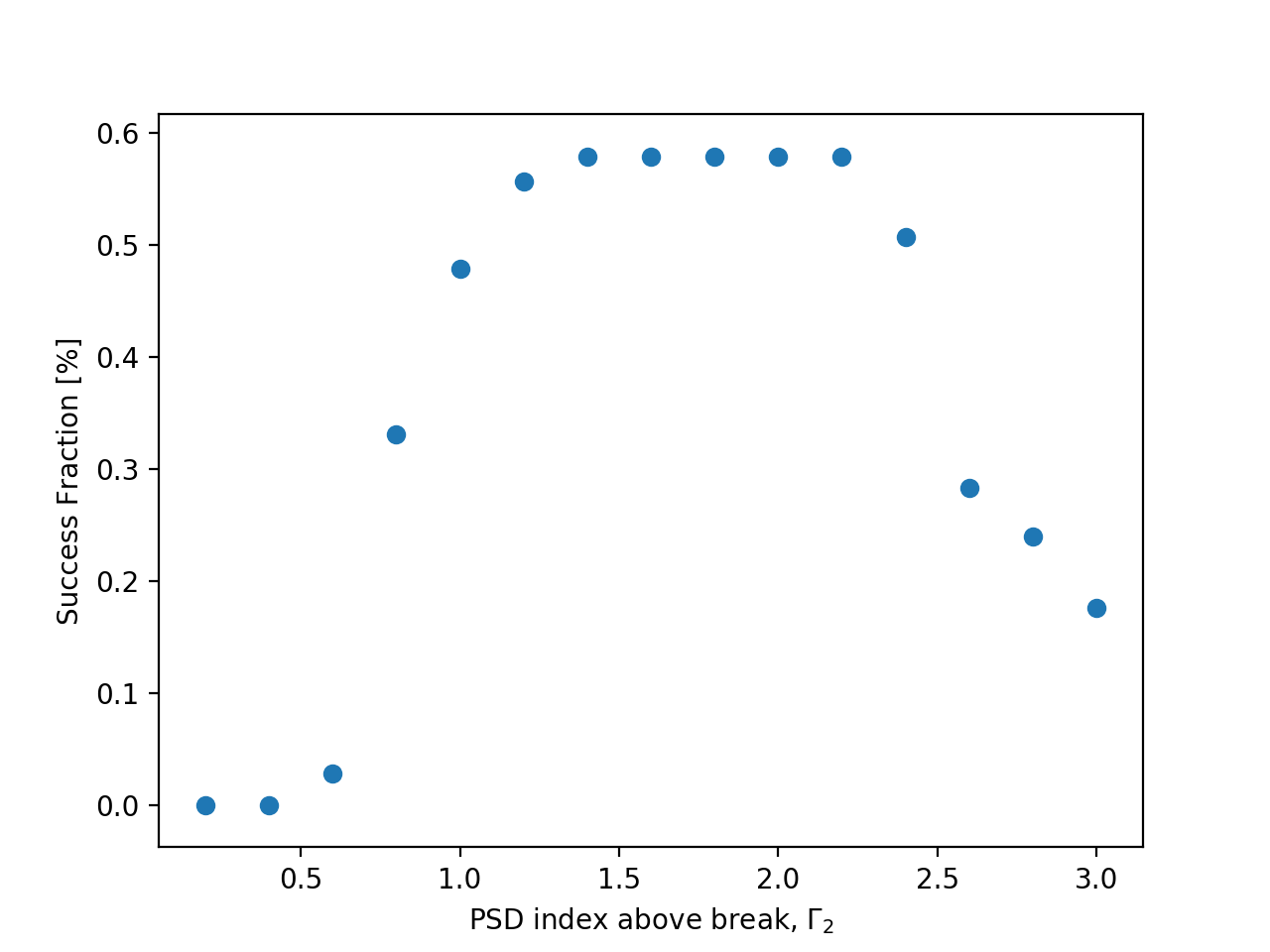}
  \includegraphics[width=0.44\textwidth]{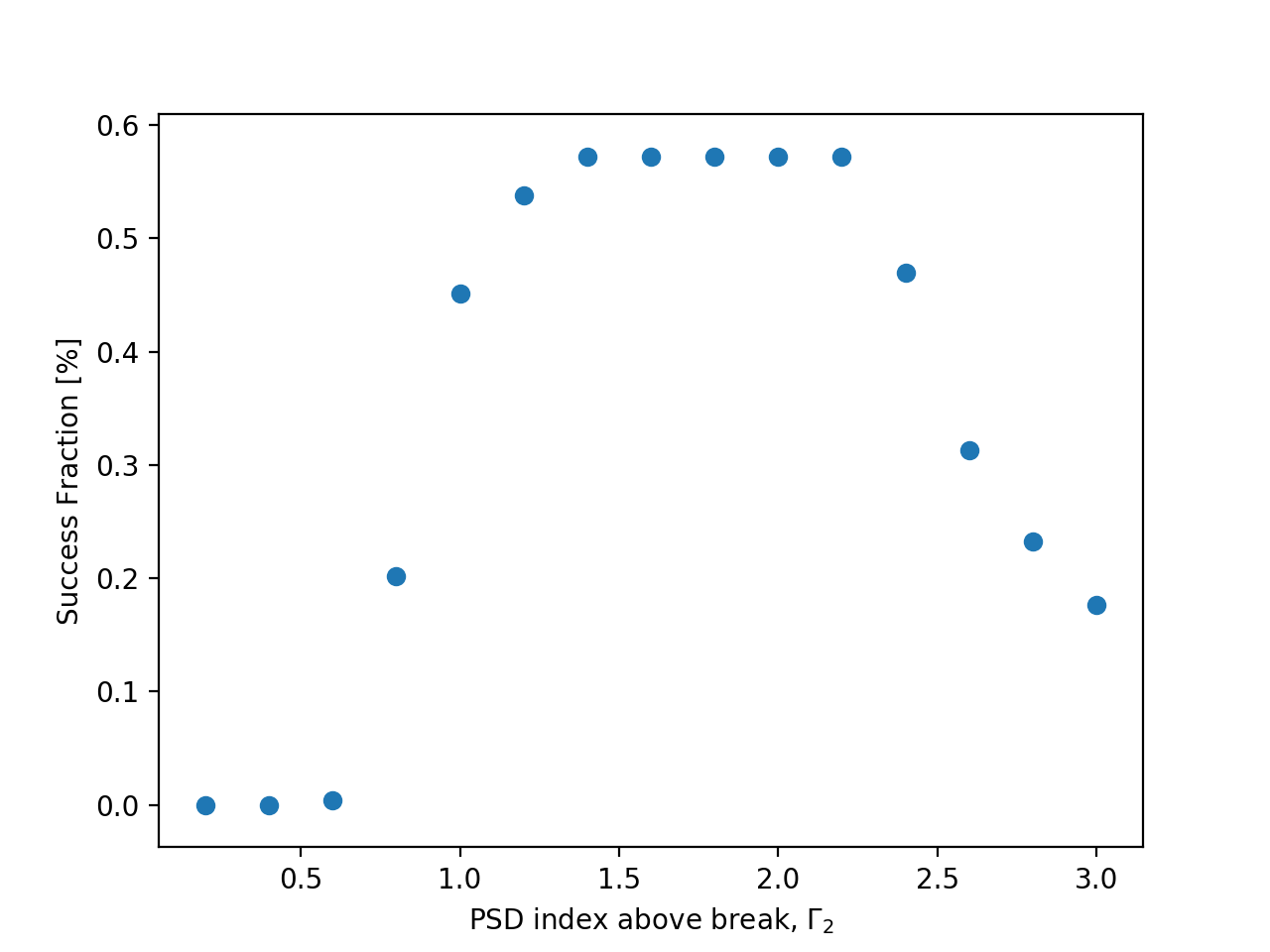}
  \caption{Red noise or $\Gamma = 2.0$}
  \label{fig:BPLvsBrPLPSD2}
\end{subfigure}
\caption{The figure shows comparison between the reconstruction with a broken power-law and bending power-law model (break at $\sim 640$ years) for a true power-law observation with {\em top}: pink noise and {\em bottom}: red noise characteristics. There is not a significant difference between the two models. However, it is clear that reconstructions are more challenging at greater values of index. }
\label{fig:BPLvsBrPLpinkred}
\end{figure}

\begin{figure}
\begin{subfigure}{.5\textwidth}
  \centering
    \includegraphics[width=0.32\textwidth]{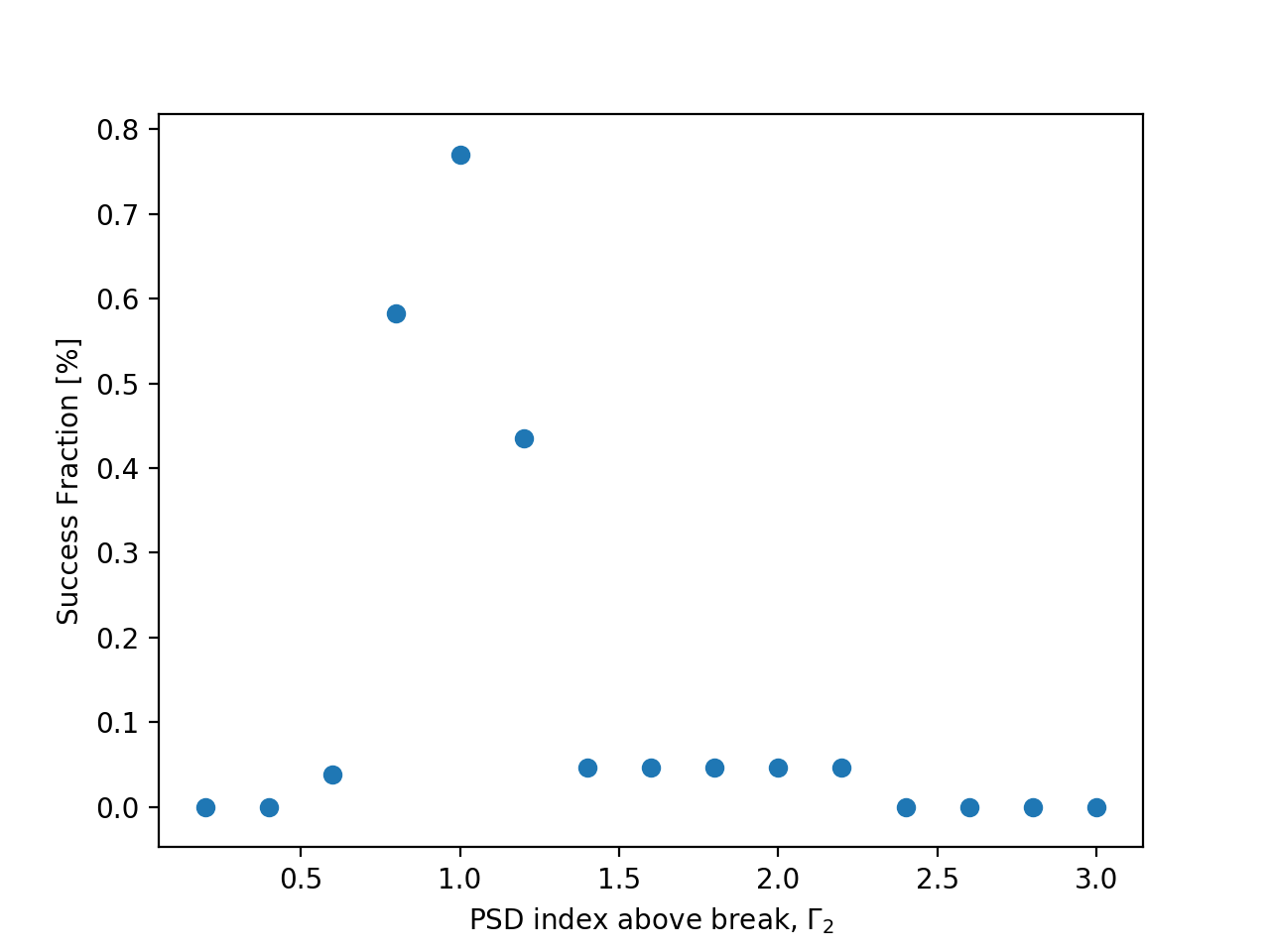}
    \includegraphics[width=0.32\textwidth]{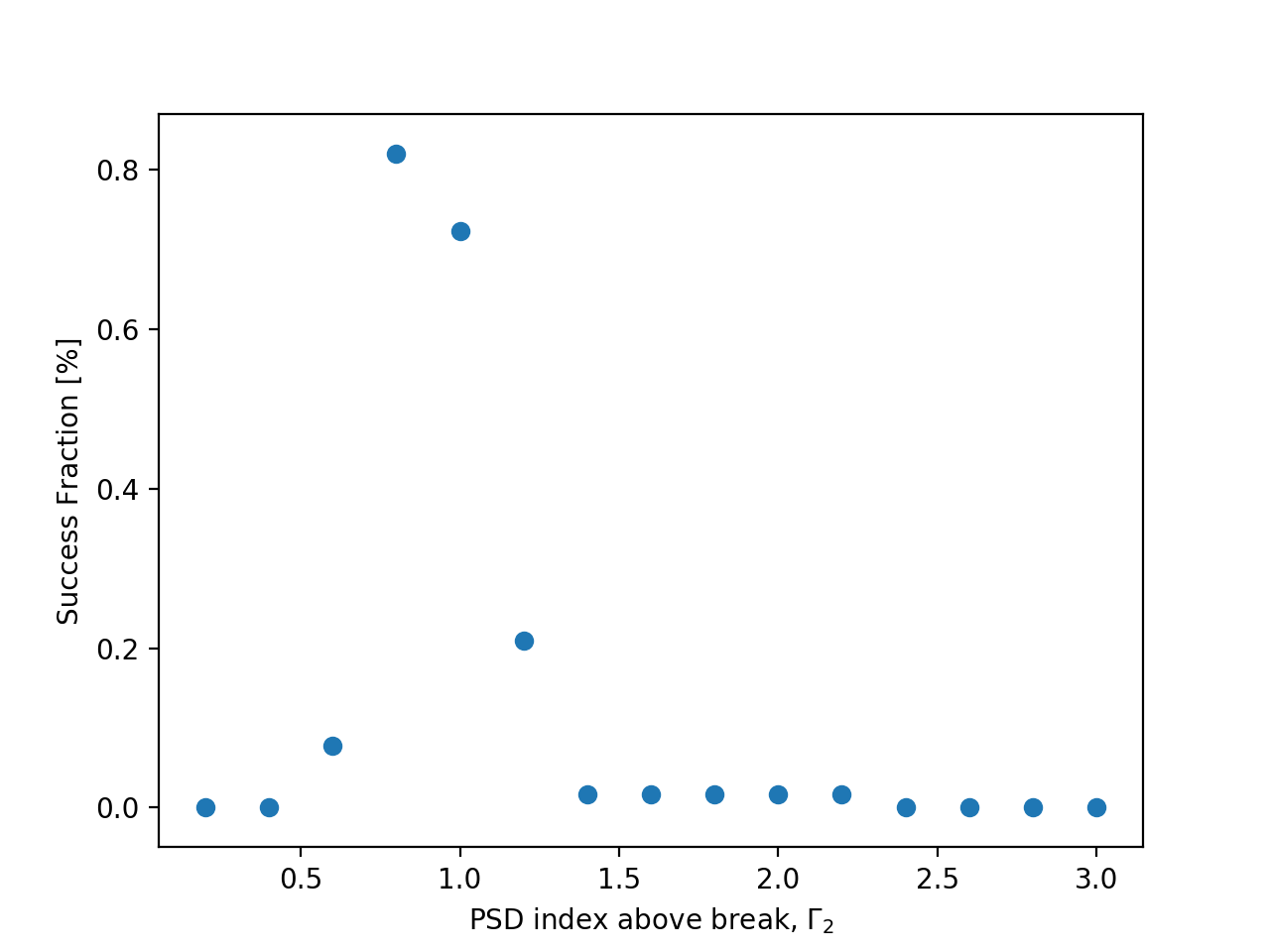}
    \includegraphics[width=0.32\textwidth]{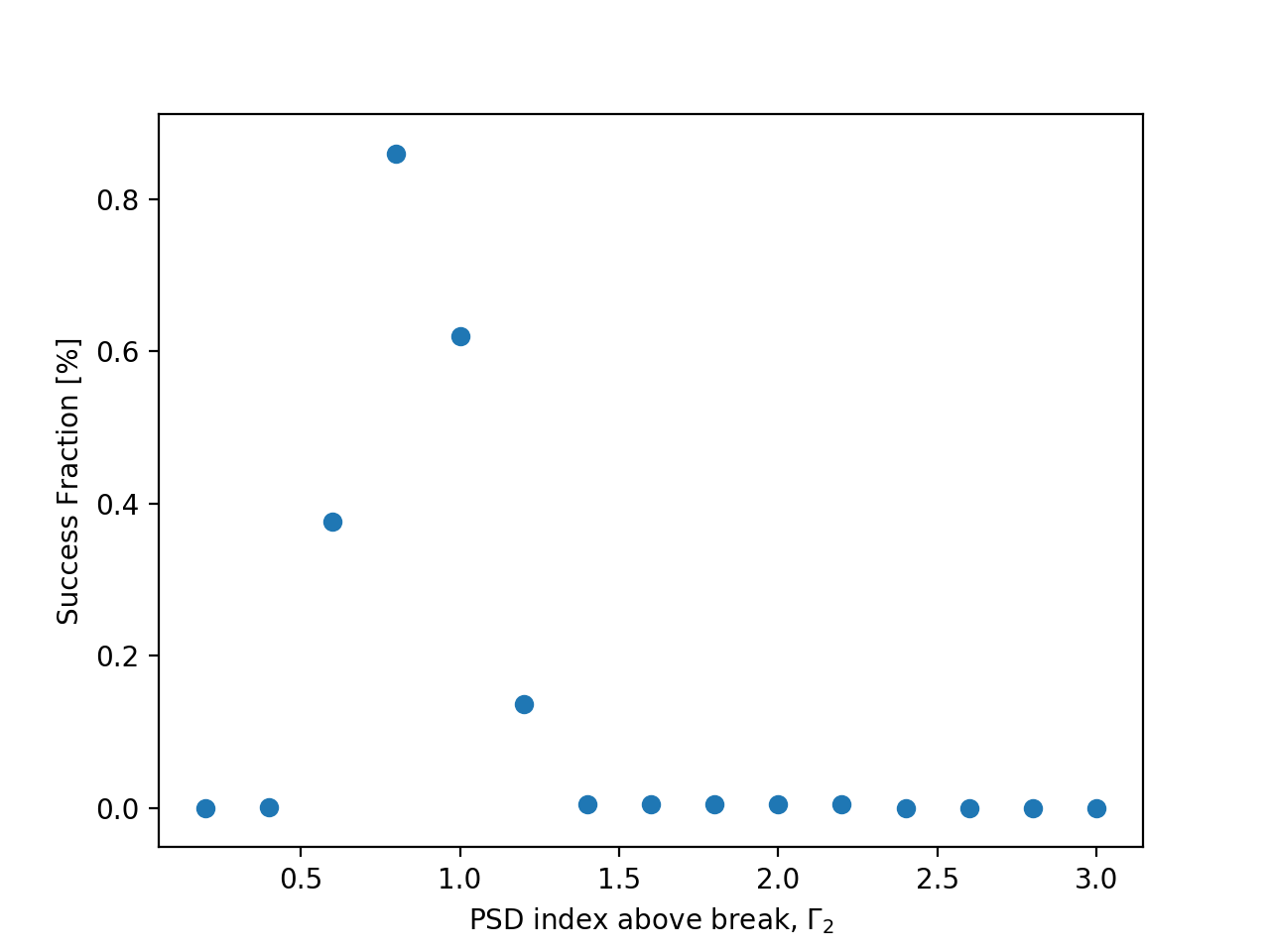}
    \caption{$\t{b} = 6.4\times10^{1}, 1.3\times10^{2}$, and $3.2\times10^{2}$}
    \label{fig:0p6501sub-first}
\end{subfigure}
\begin{subfigure}{.5\textwidth}
  \centering
    \includegraphics[width=0.32\textwidth]{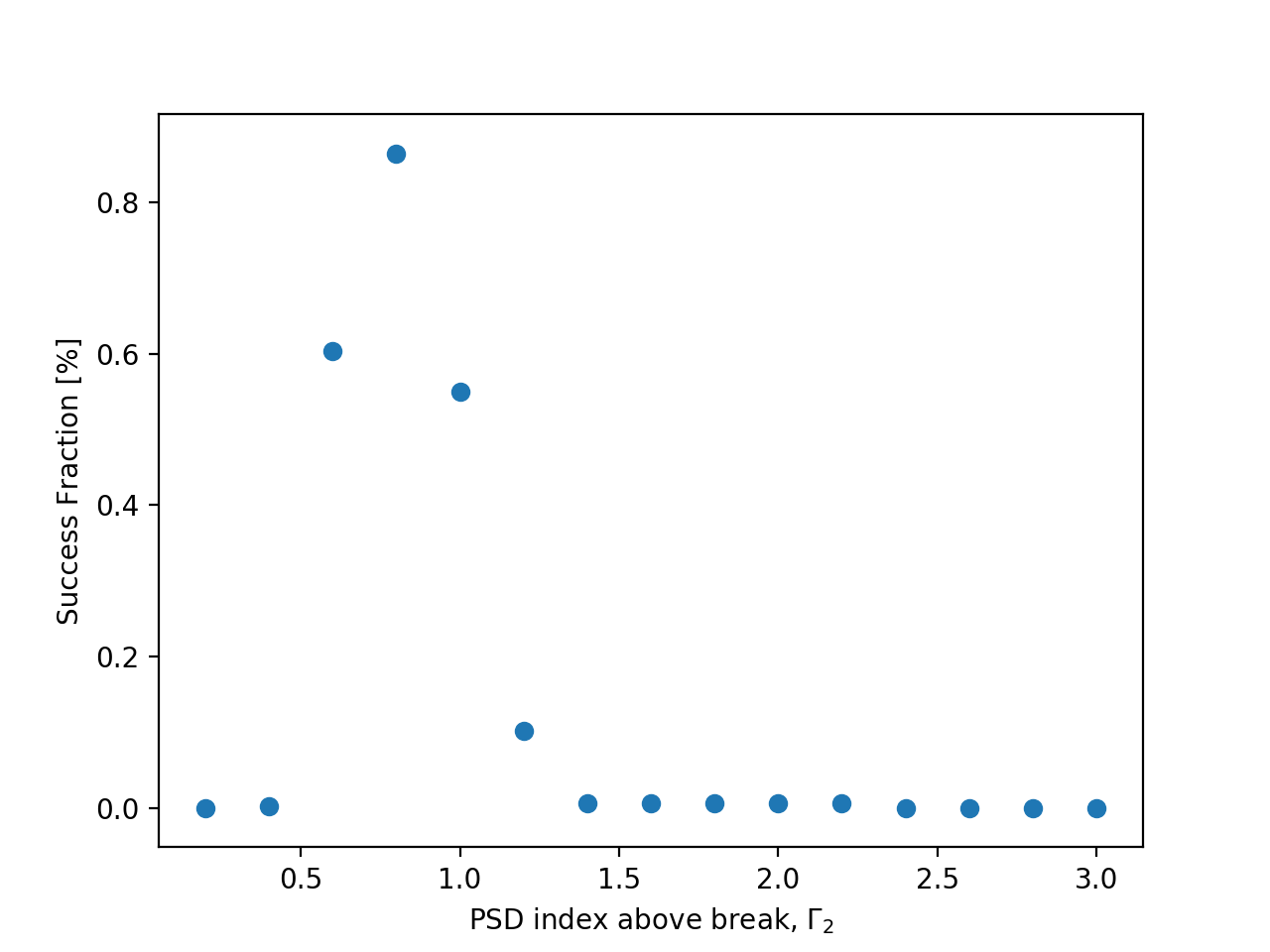}
    \includegraphics[width=0.32\textwidth]{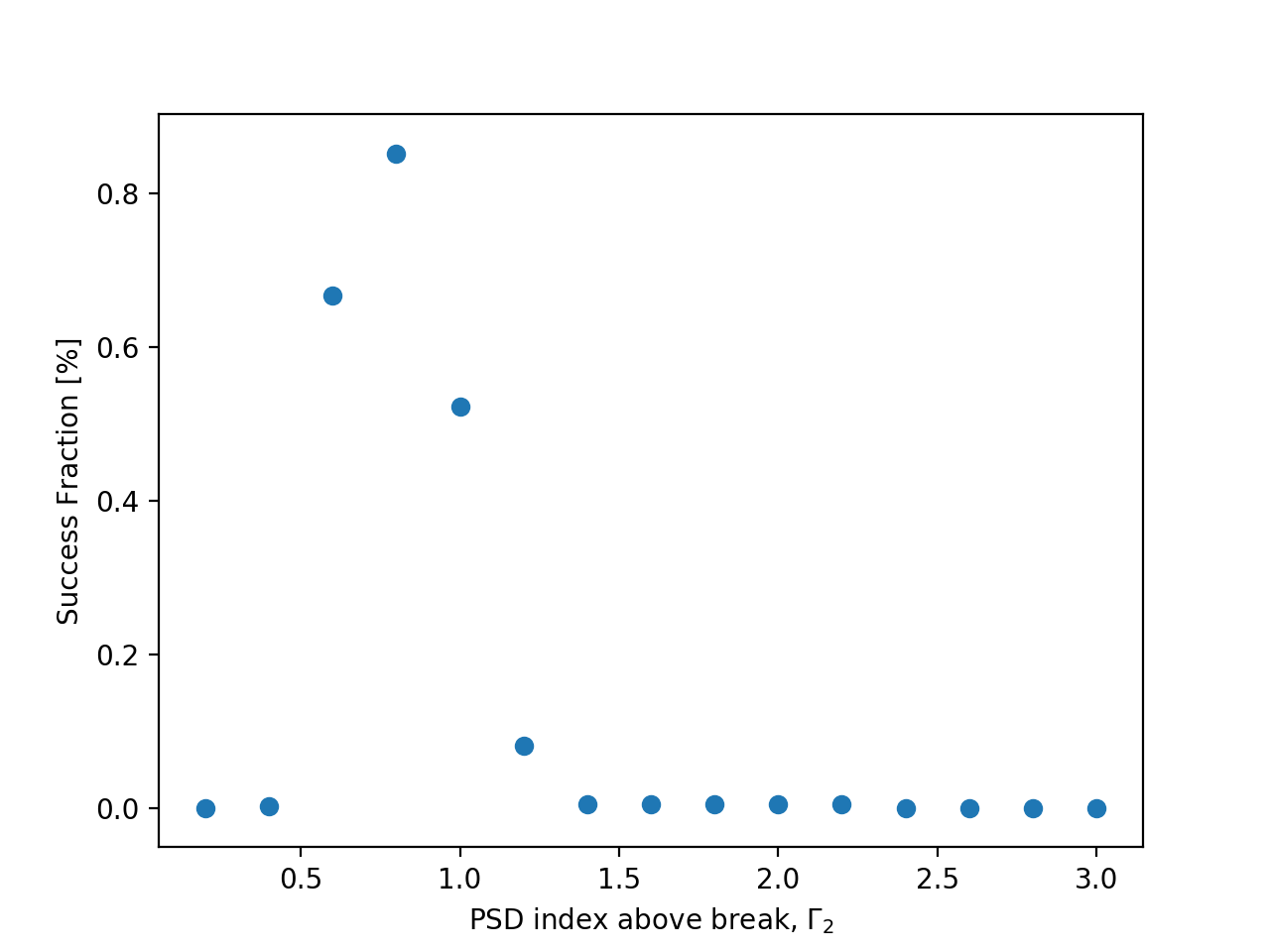}
    \includegraphics[width=0.32\textwidth]{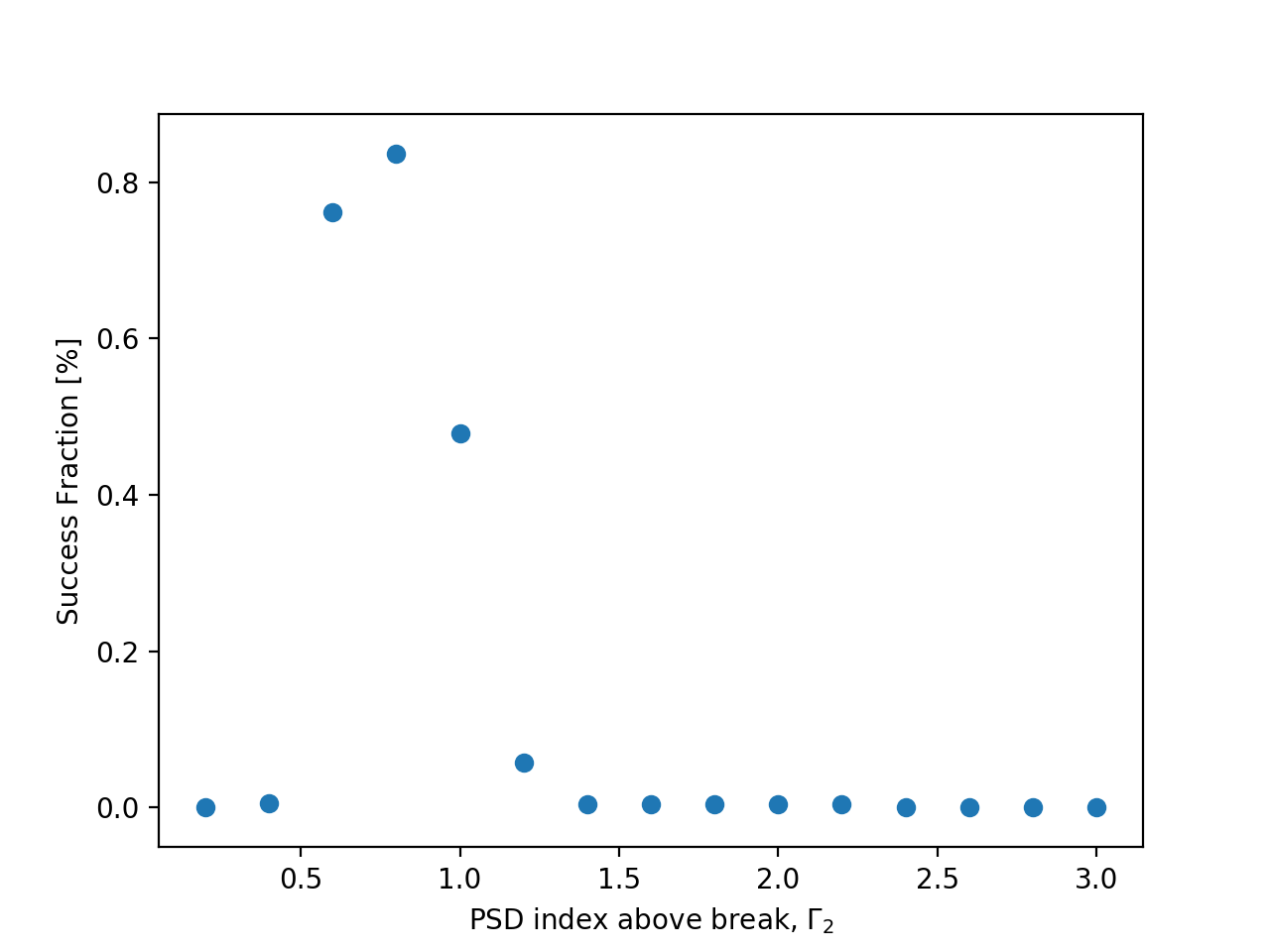}
    \caption{$\t{b} = 6.4\times10^{2}, 1.3\times10^{3}$, and $3.2\times10^{3}$}
    \label{fig:0p6501sub-second}
\end{subfigure} 
\caption{The effect of the break position on reconstructing the PSD index of simulated 
lightcurve matching the \fl cadence for Mrk 421 with a power-law with index $\Gamma =1.0$. 
Testing with break values of $6.4\times10^{1}, 1.3\times10^{2}, 3.2\times10^{2}, 6.4
\times10^{2}, 1.3\times10^{3}$, and $3.2\times10^{3}$, years respectively, we find the 
optimal value corresponding to the best reconstruction to be at $6.4\times10^{2}$ years.}
\label{fig:0p6501}
\end{figure}

Estimates of PSD indices can get pretty involved even when the true PSD is a simple,
single power-law noise. The following are some astrophysical and methodological 
issues that come up:
\begin{enumerate}
\item The flux (probability) distribution function (PDF) and PSD estimates are in 
general interconnected \citep[e.g.,][]{2003MNRAS.345.1271V, 2018Galax...6..135R,
2019MNRAS.489.2117M}. Indices steeper than one lead to deviations from strict 
stationarity and thereby Gaussianity of simulations. 
\item A transition (break or smooth bend) from coloured noise to white noise can 
prevent the divergence of total power and remedy this problem. However, the 
existence of genuine astrophysical breaks will lead to poorer estimates when 
considering pure power-law generative models.
\item Furthermore, red-noise leakage makes the power at different frequencies 
dependent; this reduces the scatter at high frequencies. This can affect the 
estimates depending upon the break or bend frequency 
\end{enumerate}

The connection between variable flux PDF and PSD estimates is an important one 
as shown in \citet{2019MNRAS.489.2117M}. For noise properties steeper than pink 
noise, there is a loss of strict stationarity. 
This results in the full PDF not being constant. In other words, higher moments 
including the variance are not conserved and therefore, the PSD estimates are
compromised. Introducing a break or a smooth bend to flatten out the spectrum at 
low frequencies or longer timescales helps to ameliorate this effect. It must be 
noted that, a priori, such a break need not necessarily be a physical break; instead, 
it is introduced beyond observed timescales in order to prevent the total power at 
low frequencies and preserve the PDF. However, as we will see in the following, 
the break that allows one to reconstruct the indices correctly, turns out to be 
compatible with the physically motivated value. 
Introducing a break at a Fourier frequency outside of the observed range works 
rather well at preserving the normal behaviour of TK lightcurves. This can be 
seen by evaluating the Gaussianity of simulated lightcurves as e.g. done for 
Mrk 421 in \citet{2020Galax...8....7C}. 
These results are in principle subject to the assumptions that the simulations 
themselves are consistent with a normal distribution as assumed in the TK approach.
Generalizations to lognormal ensembles, however, can be done by exponentiating the 
fluxes with the appropriate transformed mean and variance, which implies that the 
log of fluxes follow a normal distribution \citep{UMV:2005}. This 
presumes that the simulated TK95 lightcurve follow the Gaussian distribution 
specified at the input. Performing tests of Gaussianity on the simulations 
tests this. 
Note the vast difference between the fraction of simulations rejected for the 
PSD index 2.0 relative to index 1.0 as shown in \citet{2019MNRAS.489.2117M}. 
This shows clearly that the pure PL ensemble does not preserve Gaussianity. 
This would also mean that the lognormal simulations which are obtained by the 
exponential transform from Gaussian simulations, do not preserve the lognormal 
distribution.

\section{PSD estimation}\label{sect:estimate}
We use simulated lightcurves obtained following the description in
section~\ref{sect:method} to explore the quality of PSD reconstruction.

We simulate lightcurves for a range of PSD power-law indices from 0.2 to 3.0 for a normal 
PDF with mean and variance to match those of the \fl\ observations of Mrk 421 as an example. 
These artificial lightcurves need to match the observed mean and variance to be "realistic".
We choose a frequency range from $f_{\rm min}$ (or $T_{\rm max}$, e.g., related to the 
observation time, $T_{\rm obs}$) and $f_{\rm max}$ (or $T_{\rm min}$, i.e., related to 
the bin size, $T_{\rm bin}$).
When the simulations have a finer binning, we average simulated fluxes around the observed 
time $t_{\rm bin}$. However, in most cases we simulate the exact cadence of the observations.
The gaps would naturally be retained as gaps in order to fold in their effects. 
We then estimate the "best" PSD as prescribed in \citep{2008ApJ...689...79C}.
We compute the deviation in chi-square from within the simulations, $\chi^{2}_{\rm sim}$, 
relative to its mean. We also compute the chi-square of the simulations relative to 
observations, $\chi^{2}_{\rm obs}$. Comparing the two, we can determine the fraction of 
simulations for which the chi-square relative to observations is less than that with 
respect to simulations themselves. This is the success fraction, showing the fraction of 
cases, wherein the observations are compatible with simulations to within the dispersion 
of the simulations.

\subsection{Break position - Reconstructing index for simulated lightcurves with 
known values}
\label{sect:breakpos}
To probe if there is a universal parametrization of the position of the break that 
may be applicable to lightcurves in general, we use simulations to reconstruct the 
PSD index of a mock observation. Note that when performing such tests with simulations, 
the mock observation or lighcurve may in principle be generated by a PL, BPL or BrPL
model following eqs.~(\ref{eqn:PLmodel})-(\ref{eqn:BrPLmodel}). And if there is a physical 
break outside of the observational bandwidth of frequencies [$\f{min},\f{max}$], then 
the observation is in effect, one realisation of such a process that can be parametrised 
by a BPL or BrPL model. Therefore, in this scenario to be fully self-consistent, one 
should use BPL for both, the mock observation and the ensemble. Obviously, as we do not 
know the break value ($f_{\rm b}$) apriori, in general we would need to explore the 
reconstruction of both, the break value and the index $\Gamma$. However, this will 
require a full 2D exploration or more generally a multidimensional exploration of all 
the PSD and PDF parameters. This is beyond the scope of this paper and will be reserved 
for a future work. Here for simplicity, we run tests to see how well the index of a PL 
mock observation is reconstructed by BPL and BrPL simulations. This is illustrated by 
the example in figure~\ref{fig:BPLvsBrPLpinkred}. Clearly, reconstructions are subject 
to noise for stochastic processes generating lightcurves. Upon investigation, we find 
that the BrPL model does not necessarily have an advantage compared to the BPL model. 
The former does not guarantee a better reconstruction. On the other hand, the smoother 
BPL is a more physical parametrization. Therefore we will focus in the following on the 
BPL model. From hereon we also do not distinguish between break and bend and use them 
interchangeably. 

The BPL model of the form in eq.~(~\ref{eqn:BPLmodel}) is used to generate the ensemble 
with a range of indices $\Gamma$. 
We then compare each member with the mock observed lightcurve. We compute the reduced 
chi-square, $\bigg(\chi^2/\nu\bigg)_{\rm i}$ from the difference of simulated and 
observed periodogram for each ensemble member. We then compare with the mean reduced 
chi-square, $\bigg(\chi^2/\nu\bigg)_{\rm mean}$ that measures the dispersion within 
the ensemble. The fraction of simulations, $m$, for which the deviation of the mock 
observation from the ensemble is less than the ensemble dispersion, 
$\bigg(\chi^2/\nu\bigg)_{\rm i} < \bigg(\chi^2/\nu\bigg)_{\rm mean}$ quantifies the 
probability of the model PSD (parameter) fitting the observation. The simulated index 
with the highest fraction $m$ is the best fit index value. The absolute value of $m$ is indicative of how well a model with a certain level of complexity or number of parameters describes the features of the observed lightcurve relative to the natural statistical dispersion of the simulations with that model. 
%
%
We then repeat this for different break frequencies, $\f{break}$. 

For our artificial lightcurves, the considered estimation will reconstruct the true 
power-law PSD index the best, when the position of the break, $\f{b}$ is 
sufficiently lower than the minimum Fourier frequency of the observation window, 
$\f{min} = 1/\t{max} = 1/\t{obs}$. 
This is because if $\f{b}$ is too close to $\f{min}$, then the transition to 
white noise will reduce the power in the observed band, [$\f{min},\f{max}$] and 
produce a deviation from the true power-law. The obvious caveat is if the true 
PSD itself would have a physical break, such as related to the
viscous timescale in the disk (cf. Sec.~\ref{sect:analytictoy}) or associated 
with particle cooling \citep{fink2014}, and be close to being detected. 

%
For convenience, we employ a parameterization for the break position of the form,
\beq
\left(\frac{\f{break}}{1~{\mathrm{day}}^{-1}}\right) 
= \frac{1}{\alpha} \frac{\t{bin}}{\t{obs}}\,
\eeq
where $\alpha$ is a "lengthening" multiplicative factor. The ratio of the bin size 
$\t{bin}$ and the length of the observation window, $\t{obs}$, represents the dynamic 
range and is a natural scale to consider. Hence, $\alpha$ essentially parameterizes, 
how many times this dimensionless scale the frequency break is placed at.
For the considered {\it{Fermi}}-LAT observations of Mrk 421 with a duration of about 
6.5 years and a bin size of 10 days \citep{421logN}, $\t{bin}/\t{obs} \simeq 4.2 
\times 10^{-3}$. 
Therefore, a factor of e.g., $\alpha = 1000$ would correspond to a break frequency, 
$\f{b} \simeq 4.2\times10^{-6} \dinv$, corresponding to a timescale of $t_b 
\simeq 640$ years. 

Our simulated lightcurves each have a length of 2340 days, corresponding to the length 
of the \fl observations of Mrk 421, with the same bin size of 10 days. 

The factor, $\alpha$ is set to values 100, 200, 300, 500, 1000 and 3000 which
corresponds to break or bending timescales of $t_b = 6.4\times10^{1}$,
$1.3\times10^{2}$, $1.9\times10^{2}$, $3.2\times10^{2}$, $6.4\times10^{2}$ 
and $1.9\times10^{3}$ years, respectively.
We then perform estimation of the index of these "mock observed lightcurves" 
with the methodology described above. Naturally, using artificial lightcurves, 
we expect the estimates to match the chosen index value, e.g., $1.0$, used to 
generate the lightcurves in the first place. As shown in Figure~\ref{fig:0p6501},
the value of $\alpha = 1000$ corresponding to $t_b = 6.4\times10^{2}$ yrs 
gives the sharpest reconstruction of the correct index. The predictions work
better with index $\lesssim 1$, though even for indices  $> 1$, the predictions
seem to be much better than when affected by non-stationarity. 
%
We also ran a few trial runs with the DE13 method. From these, it was unclear
that there is any advantage vis-a-vis precision for the available data 
especially for indices greater than 1.0, and given the computational cost of 
the DE13, the bending power-law seems advantageous especially for simpler PSD
models. The effects of a more complex input model of PSD and PDF needs a
dedicated paper and is left for the future. Thus, given the advantage in
terms of accuracy and precision of DE13 is not clear for the longest 
timescales and the computational cost is high, the use of a BPL model that 
is faster and easier to interpret is advantageous.

\begin{figure}
  \centering
    \begin{subfigure}{.5\textwidth}
    \centering
       \includegraphics[width=0.9\linewidth]{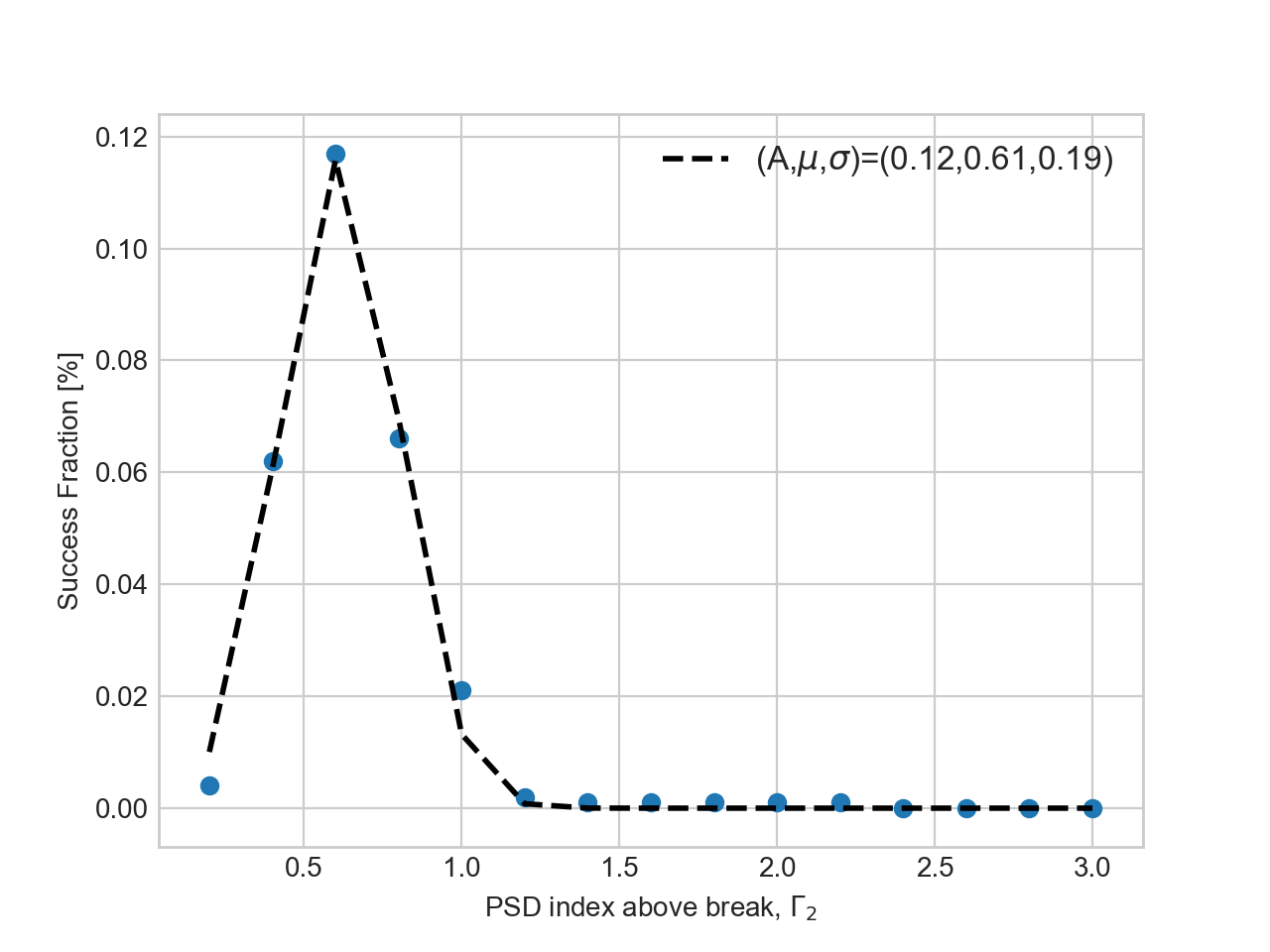}
    \end{subfigure}
    \begin{subfigure}{.5\textwidth}
    \centering
    \includegraphics[width=\linewidth]{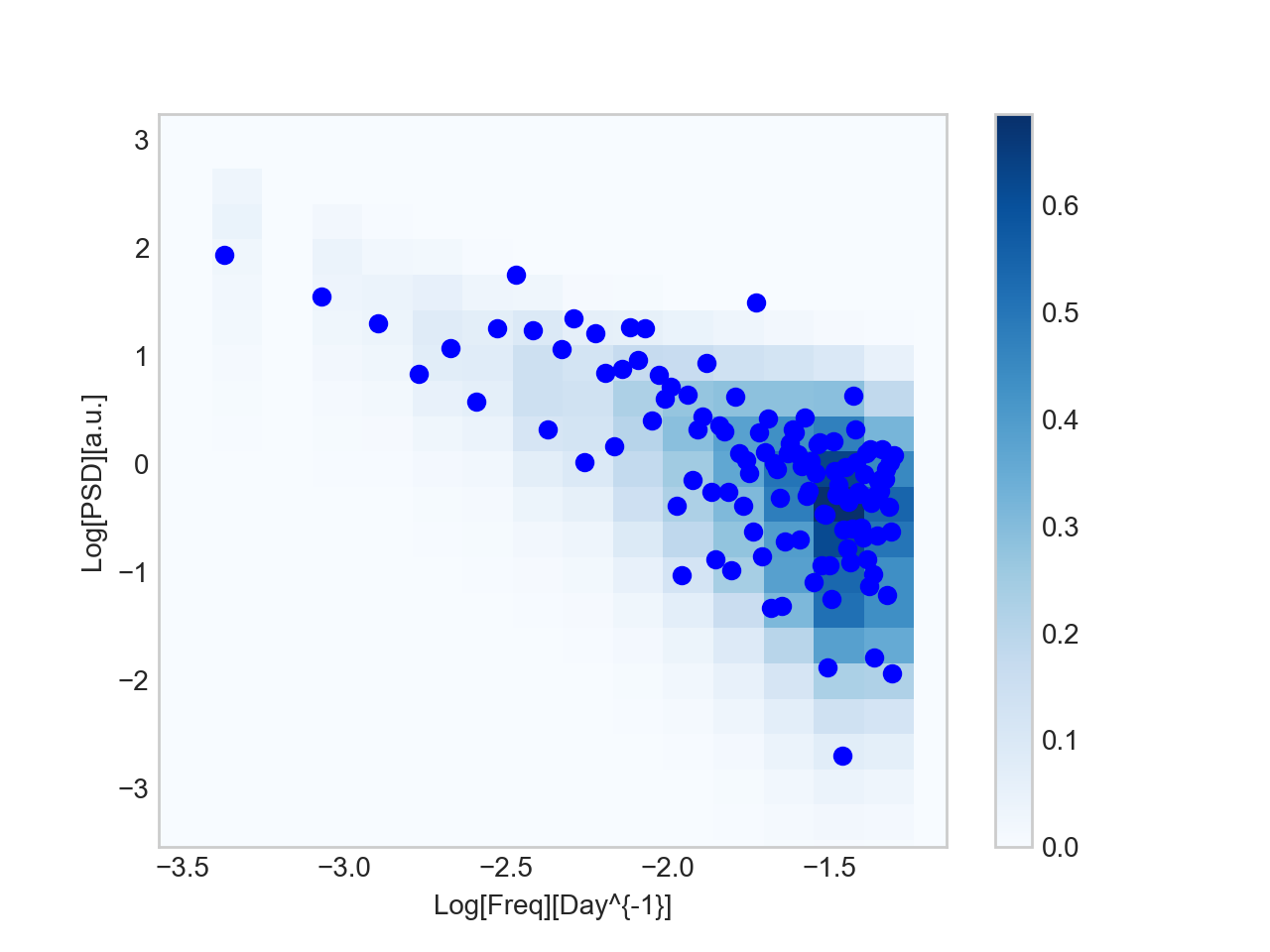}
    \end{subfigure}
\caption{The PSD index of the \fl lightcurve determined with a break factor of 
1000 or $\t{b} \approx 640 $ years. {\em Top}: From the chi-square test, the 
maximum number of simulations prefer an index of value $\Gamma \approx 0.6$. 
Note that this does not account for a bias which from our tests is $\approx 
0.2$. {\em Bottom}: The probability map quantifying number of simulations that 
have a certain power density in a given frequency bin is shown. The blue circles 
are the periodogram directly obtained from the observed \fl lightcurves. The color bar represents the fraction of simulations in any given bin having been normalised to the total number of simulations.}
\label{fig:fermipsd421}
\end{figure}

\subsection{PSD of LAT and BAT lightcurve}
\begin{figure}
  \centering
  \begin{subfigure}{.5\textwidth}
  \centering
    \includegraphics[width=0.9\linewidth]{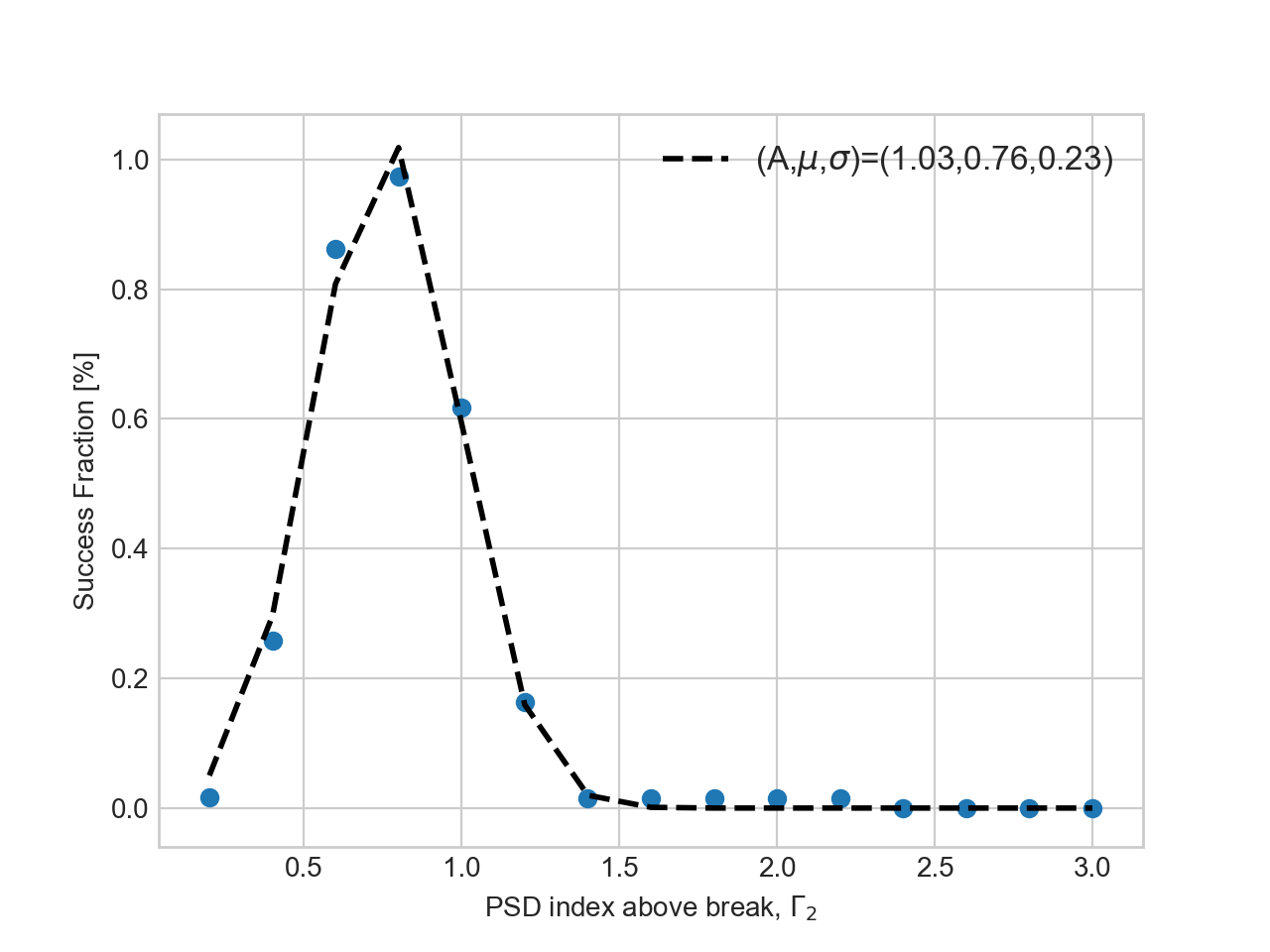}
    \end{subfigure}
    \begin{subfigure}{.5\textwidth}
    \centering
    \includegraphics[width=\linewidth]{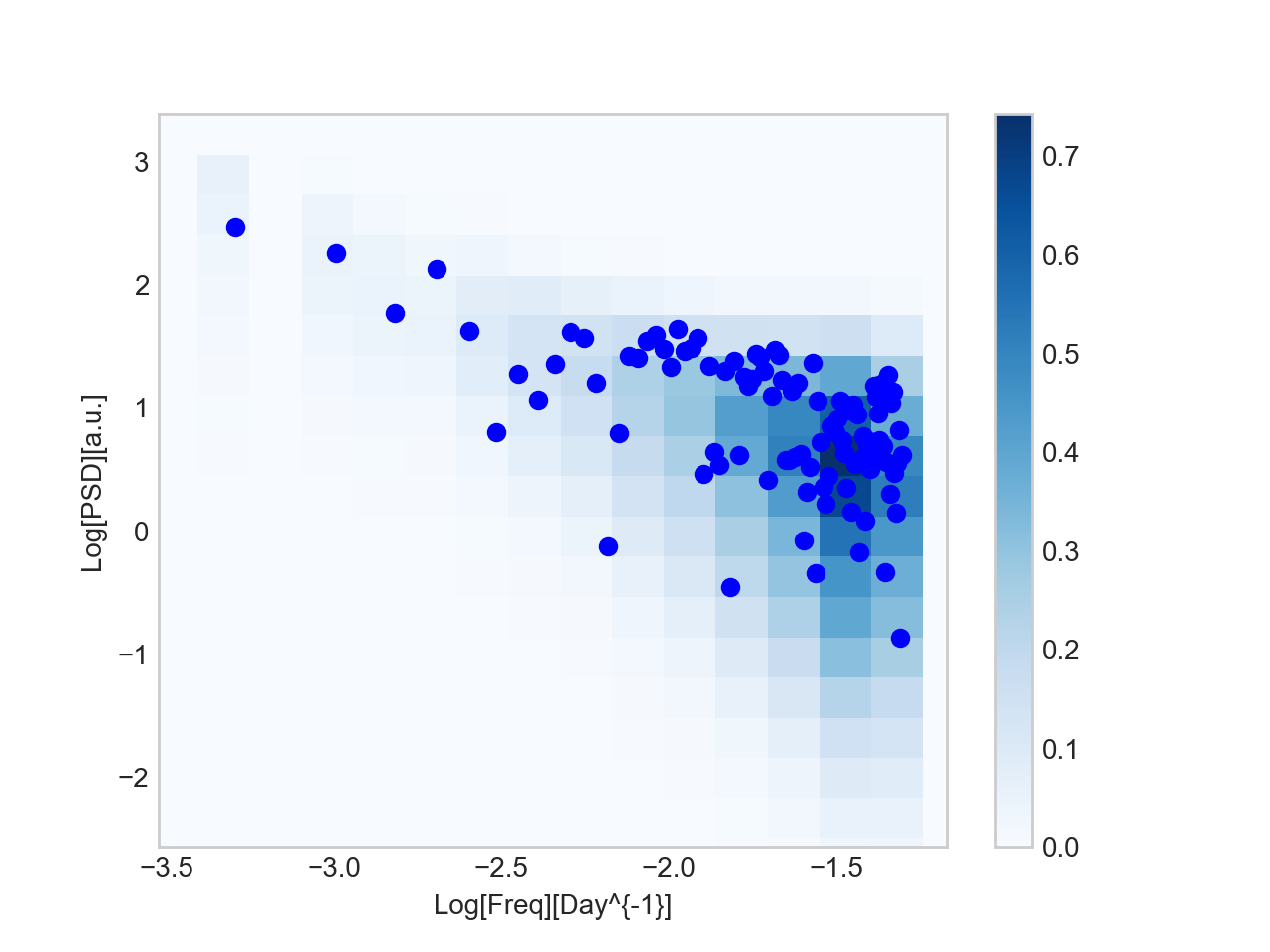}
   \end{subfigure}
\caption{Figure shows the PSD index of the Swift-BAT lightcurve determined with 
a break factor of 1000 or $t_b \approx 640$ years assuming a Lognormal PDF. 
{\em Top}: From the chi-square test, the maximum number of simulations prefer 
an index of value $\Gamma \approx 0.8$. Note that this does not account for a 
bias which from the tests earlier is $\approx 0.2$. {\em Bottom}: The simulated 
probability map for Swift-BAT lightcurves as well as the periodogram of the 
observation as in fig.~(\ref{fig:fermipsd421}).}
\label{fig:batpsd421}
\end{figure}

Having tested on mock longterm lightcurves of timescales of a few years, we 
now evaluate the PSD of \fl and Swift-BAT lightcurves of Mrk 421. The length 
of both of these lightcurves is comparable at $\sim 2000$ days or $\sim 7$ 
years. Furthermore, the binning is at around 10 days. We therefore, pick the
"optimal" simulation break factor determined from our tests at $\t{b} \approx 
640 $ years or break factor, $\alpha = 1000$. The analysis in
\citet{2020Galax...8....7C} indicates that the preferred distribution for 
the \fl lightcurves is lognormal and even for Swift-BAT, the lognormal PDF
matches the high flux tail. Hence, for the simulated ensemble we use the
lognormal function as input PDF. With these, we perform the reduced 
chi-square test \citep{2008ApJ...689...79C}, scanning index values in the 
We also provide the probability map - the two-dimensional histogram of power 
spectral density of the simulated ensemble with these indices, stacked upon 
each other and the periodogram of the lightcurves. The top panels of 
figures~(\ref{fig:fermipsd421}) and ~(\ref{fig:batpsd421}) show the success 
fraction as a function of index and the bottom panels show the PSD map for 
the \fl and Swift-BAT lightcurves, respectively. In both cases, the maps 
show that the scatter of the PSD for observed cadence (10 days) is at the 
level of the average dispersion of the simulations. This shows that the 
good quality of monitoring with \fl and Swift-BAT allows for a relatively 
precise estimate of the long-term stochastic properties. The peak of $m$ for the \fl case is well defined but has a value of only $\sim 12 \%$ similar to \fl and OVRO results in \citet{2020MNRAS.494.3432G}. This suggests that of the BPL or "PL with smooth break/bend models", the "best" choice is index $\sim 0.6$. This is a sense a reduced BPL model as we fix the lower index (below break) to zero, and while determining best higher index (above break) we fix the break frequency to the optimal one obtained from simulations for the cadence. The low $m$ values indicate, these simple models are not sufficient to explain the observations and we need to explore more complex models. The natural next step would be to jointly explore the space of break and the index above it, as we have already indicated in section~(\ref{sect:breakpos}). This reduced BPL model seems to work much better for Swift-BAT lightcurves. This is consistent with the probability or PSD maps. The BAT maps and blue data points in figure~\ref{fig:batpsd421} appear to have more of a mean PL type behaviour about which there is scatter and the best $\Gm$ describing this has a central value of 0.8. Relatively, the LAT maps have a slight hint of curvature. However, a definitive conclusion on this would depend upon full multi-dimensional parameter estimation.   


As mentioned, we find that for the \fl lightcurve, the index obtained from the reduced chi-square 
test is $\Gamma \approx 0.6$ whereas for the Swift-BAT lightcurve we obtain $\Gamma 
\approx 0.8$.
The uncertainty quantification derives from parameter estimation by varying the 
relevant model parameters (in this case, the PSD index) in an appropriately wide 
range and freezing other parameters \citep[as e.g. done in][]{2020MNRAS.494.3432G}. 
Estimates of the statistical uncertainty are made here by fitting a Gaussian to 
the success fraction, giving $0.2$ for both lightcurves. Neither of these values 
account for a systematic bias as inferred from our methodological 
tests using mock observations with known index, whose value is $\sim 0.2$. This
value denotes the difference between the reconstructed index value where the 
success fraction peaks and the true value of the index for simulated or mock 
observations. 
Accordingly, the LAT and BAT lightcurves are characterized by PSDs with an index 
above the break, $\Gm = 0.6\pm0.2^{stat}+0.2^{sys}$ and 
$\Gm = 0.8\pm0.2^{stat}+0.2^{sys}$, respectively. 


Our findings for Mrk 421 can be compared with related results in the literature. 
\citet{2020MNRAS.494.3432G}, for example, report PSD indices of $1.1\pm0.4$, $1.3
\pm0.7$ and $1.1\pm1.6$ for \fl, Swift-XRT (0.3-10 keV) and RXTE-PCA (3-20 keV) 
lightcurves, respectively. The \fl result is compatible, within errors, 
with pink noise, while their X-ray estimate is not well-constrained.
\citet{2018ApJ...859L..21C}, on the other hand, find that in the X-rays the 
extended, short- to long-term PSD is compatible with a bending PL model 
(breaking from $\Gamma \simeq 1.2 \pm0.5$ to $\Gamma  \simeq 2.5\pm0.5$ 
at a high-frequency break corresponding to $\sim 10$ d). While this indicates
that the extended PSD can show further bends or breaks, the longterm PSD 
estimate ($1.2\pm0.5$) is again compatible with pink noise. This compares 
well with our findings reported here. For the transition between different
regimes, especially at shorter timescales, an extended examination of the 
PSD and break at timescales of days would be needed, which is not the focus 
of this work. Note that the break frequency we employ corresponds to a break 
time of about $T_b \simeq 640$ yr. This is close to the viscous timescale at 
the disk truncation radius, considering a characteristic black hole mass in 
Mrk 421 of $M_{\rm BH} \sim 2 \times 10^8 M_{\odot}$ \citep[e.g.,][]{bar2003}.

\section{Discussion and Conclusions}
\label{sect:conc}
Quantifying the power spectral density is the key to understanding the stochastic 
behaviour of a variable source such as an AGN or XRB. Features such as breaks can 
be indicators of physical timescales and therefore processes and transitions between 
them \citep{fink2014, rie2019}. Knowing the PSD is also important in order to test the true 
significance of a genuine quasi-periodic oscillation against false alarms from noisy 
variability \citep{ait2020, vau2016}. Several methods have been developed to quantifying the 
PSD \citep{TK:95, 2013MNRAS.433..907E}. Central to estimating the PSD is the implicit 
assumption, in most cases, of weak non-stationarity. It has been shown that this is 
not guaranteed and is crucially compromised at larger values of PSD index 
\citep{2003MNRAS.345.1271V, 2019MNRAS.489.2117M}. This is due to divergence of the 
variance (second moment of the underlying distribution) or integral power of the 
variability in the limit of infinite bandwidth as the index exceeds values $1-1.5$. 
In practice, for large enough bandwidths such as in the case of long term monitoring 
discussed here ($[\t{min},\t{max}] =$ [days,years]), this effect is already pronounced 
as shown in \citep{2019MNRAS.489.2117M}. The "redder" the random noise in 
variability, the lower is the fraction of realisations of the process which preserve 
same probability density function (PDF). Therefore, the statistical estimation of the 
PSD index and  other stochastic properties dependent on it, is compromised for a range of values 
$(\gtrsim 1)$. Here in this work, we seek to find a way to assuage this problem. A 
bending power-law (BPL) PSD model inspired by OU type processes is used. The smooth 
transition to white noise from coloured noise prevents this divergence of power and 
is also a realistic model of high energy variability over a range of timescales from 
weeks to many years. In principle, we expect a characteristic low-frequency bend or 
break corresponding to the viscosity timescale at the disk truncation radius, 
$t_{\rm visc}$, which in turn depends upon the mass of the central supermassive black 
hole. Typical values range from hundreds to thousands of years. We find that this 
timescale is a reasonable guess for a methodological break timescale, $\t{b}$, which 
prevents the divergence and provides a good reconstruction of index values for 
variability over weeks to years. Additionally, by virtue of being a simple modification 
of the PSD model as opposed to a fully general model of the variability (with multiple 
free parameters for the PSD and PDF), it is computationally very fast. This method is 
applied to X-ray (Swift-BAT) and gamma-ray (\fl) long term lightcurves of the source 
Mrk 421 spanning nearly 7 years. The reconstructed PSD values for both 
lightcurves are in principle compatible with pink (flicker) noise, given the uncertainty 
and bias in the estimate, though there seems to be a tendency for a slightly flatter 
value at gamma-ray energies. These findings are broadly compatible with previous results \citep{2018ApJ...859L..21C,2020MNRAS.494.3432G}. 
Our evaluation with the BPL model also provides realistic estimates of statistical and 
systematic uncertainty (each $\lesssim 25\%$) with the success fraction from reduced 
chi-square \citep{2008ApJ...689...79C}. This makes the estimate a rather robust one. 
We note that for a full reconstruction of variability parameters, we cannot avoid a 
multi-dimensional parameter reconstruction. This will be the subject of a future 
investigation. 
In general, quantifying the variability properties of a source is inherently, 
multi-observable problem. The PSD quantifies the underlying stochasticity and the 
PDF encodes the form of the driving mechanism. As found in \citet{2019MNRAS.489.2117M}, 
these estimates are interlinked. However, making certain reasonable assumptions about 
one, one can investigate the other. In this way, any limitations of the estimate are 
made explicit and reliability is proportional to the quality of the data. For getting 
such an estimate of the index for a simple model (e.g., power-law regime) of 
stochasticity in the long-term variations, the use of a BPL model with a bend 
frequency comparable to viscous timescale at the disk truncation radius, $f_b \sim 
1/\t{b}$, is a physically motivated, efficient and accurate solution. 

\section*{Acknowledgements}
We thank Atreyee Sinha for providing us with the observational 
data for Mkn 421.  NC thanks the European Research Council funding via the CUNDA project under number 694509. NC also kindly acknowledges the support from MPIK, Heidelberg and DARC, Reading for their resources in the development of this research. FMR acknowledges funding by a DFG Heisenberg 
Fellowship under RI 1187/6-1.

\section{Data Availability Statement}
The data underlying this article were provided by Atreyee Sinha by permission. Data will be shared on request to the corresponding author with her permission.
\bibliographystyle{unsrtnat}
\bibliography{references}  






\end{document}